# *Not someone, but something: Rethinking trust in the age of medical AI*

Jan Beger*

**ABSTRACT**

*As artificial intelligence (AI) becomes embedded in healthcare, trust in medical decision-making is changing fast. Nowhere is this shift more visible than in radiology, where AI tools are increasingly embedded across the imaging workflow—from scheduling and acquisition to interpretation, reporting, and communication with referrers and patients. This opinion paper argues that trust in AI isn't a simple transfer from humans to machines—it's a dynamic, evolving relationship that must be built and maintained. Rather than debating whether AI belongs in medicine, it asks: what kind of trust must AI earn, and how? Drawing from philosophy, bioethics, and system design, it explores the key differences between human trust and machine reliability—emphasizing transparency, accountability, and alignment with the values of good care. It argues that trust in AI shouldn't be built on mimicking empathy or intuition, but on thoughtful design, responsible deployment, and clear moral responsibility. The goal is a balanced view—one that avoids blind optimism and reflexive fear. Trust in AI must be treated not as a given, but as something to be earned over time.*

## 1. INTRODUCTION

Trust is the quiet currency of healthcare. It shapes clinical judgment, anchors patient vulnerability, and holds the system together—even in times of uncertainty. Trust allows individuals to share fears with a stranger in a white coat, consent to procedures they may not fully understand, and believe in healing before the outcome is known.

But healthcare is changing. Artificial intelligence (AI) is no longer experimental; it is already diagnosing, predicting, tailoring treatments, and engaging patients. In radiology, especially, AI is now embedded across the entire workflow—from scheduling and acquisition to triage, image interpretation, report generation, and communication with referrers and patients. It brings speed and precision—but it also raises a deeper question: can we trust machines the way we trust people? Should we?

Scholars have echoed this dilemma, noting that despite numerous frameworks for "trustworthy AI," there remains no shared understanding of what trust in AI actually means in clinical care.[1]

AI holds significant potential to improve healthcare—enhancing outcomes, reducing inefficiencies, and expanding access. Yet its impact will not be measured by technical milestones alone. It will depend on how these systems are integrated into care, how they are governed, and how they earn—and sustain—trust.

This paper explores what trustworthy AI should look like in medicine. How do we balance automation with oversight? What safeguards are essential? These are not hypothetical questions—AI is already part of clinical decision-making, and its role is expanding rapidly.

---

* https://www.linkedin.com/in/janbeger/

Drawing on philosophy, bioethics, and system design, I contrast the emotional trust placed in human caregivers with the structural trust required of AI systems. Trust cannot be simply transferred from clinician to algorithm. It must be redefined. AI may not offer empathy, but it can be designed to earn trust—if we build it with care.

## 2. THE NATURE OF TRUST: HUMAN VS. MACHINE

Trust in healthcare is, at its core, a relational experience—an act of opening oneself to another with the belief, or at least the hope, that they will respond with care. This trust feels deeply human: we trust clinicians not only for their training and licensure, but because they listen, observe, and share in the weight of uncertainty. It is a kind of moral intimacy, built gradually through presence, gestures, and tone.

Artificial intelligence is changing that equation. No longer a distant possibility, AI is already embedded in clinical workflows—particularly in radiology, where scan optimization, image analysis, triage tools, and report generation are actively deployed. These are not speculative tools. They are in routine use in many departments—reducing scan times, improving image quality, flagging pulmonary nodules, prioritizing stroke cases, and even drafting structured reports. Each use case brings its own kind of trust challenge.

Unlike human relationships, where trust is shaped by emotion, ethics, and shared experience, AI does not feel, care, or understand. Yet we are increasingly asked to trust its outputs—sometimes even over the judgment of a human clinician.

That trust often relies less on deep understanding than on workflow fit: does the AI save time, integrate smoothly, and perform reliably? In radiology, qualitative studies suggest that seamless integration and speed often matter more to users than explainability—especially when tools support routine tasks and reduce workload—while voicing concerns about legal liability and silent system updates.[2] This marks a shift from emotional trust to practical reliance.

Still, not all trust is equal. A triage tool that misclassifies a stroke scan is far more consequential than a report template that mislabels a finding. Radiologists—and the systems around them—must calibrate trust accordingly. That calibration isn't always visible. In some settings, AI tools update silently or behave inconsistently across modalities, leaving radiologists unsure whether the tool they trusted yesterday is the same one they're using today.

This raises a fundamental question: how do we trust something that does not understand what it means to be trusted?

Interestingly, studies of smart healthcare adoption suggest trust often stems not from deep understanding, but from how users experience features like personalization, ease of use, and even anthropomorphism.[3] These design cues shape emotional and cognitive responses—even when the system itself remains opaque.

In radiology, that opacity is often reinforced by limited evidence. A recent review of over 170 CE-marked AI tools found that most were supported only by technical performance metrics, with few demonstrating impact on clinical decision-making or patient outcomes.[4] This weakens the foundation for confident delegation—and raises new questions about what it means to trust a tool in practice.

In practice, trust in AI is not about empathy. It is about confidence in design—how users perceive the system's competence, contextual sensitivity, and alignment with patient interests.[5]

Do we believe the model has been trained on appropriate data? That its limitations are known? That its recommendations are safe?

These aren't the questions we ask of human caregivers, but they are no less important.

The nature of trust is shifting. As AI becomes more involved in imaging and diagnosis, we must redefine what it means to trust—not by making AI more human, but by ensuring it is transparent, accountable, and worthy of delegation.

## 3. VULNERABILITY AND DELEGATION

Every encounter with the healthcare system—whether as a patient or caregiver—requires a quiet surrender. To seek care is to be vulnerable. It means giving someone access not just to the body, but to uncertainty, fear, and sometimes pain. Trust makes that surrender possible. We believe the clinician in front of us not only knows what they are doing, but will use that knowledge responsibly, in our best interest.

At its core, trust in healthcare is about delegation. We hand over some measure of control, trusting it will be used wisely. This delegation is shaped not just by clinical skill or institutional authority, but by relational trust. Research shows that higher trust is directly linked to preventive behaviors like vaccination and regular check-ups—independent of access or income.[6]

In traditional care, we delegate because we believe our clinicians see the whole picture—not just data, but context. They recognize that we are more than a diagnosis. And in return for our trust, they carry a personal, ethical, and professional responsibility. This mutual awareness of vulnerability turns delegation into a shared endeavor—not just a transaction.

When AI enters the equation, that relationship shifts. We are no longer delegating to a person, but to a system—a tool that, no matter how advanced, cannot recognize vulnerability. There is no shared gaze. No subtle reassurance. No ability to adjust to our tone, pause, or pain. At least, not yet.

Delegating to AI feels less like entrusting a partner and more like placing a calculated bet: that the system is well-designed, well-tested, and appropriately integrated. And initial trust plays a central role here. In one large-scale study, physicians' willingness to use AI systems was driven more by initial trust than by performance expectations or usability.[7]

And yet, even well-designed systems can mislead. In a clinical study, healthcare professionals often followed incorrect AI recommendations—even in high-risk cases—due to automation bias.[8] When delegation lacks shared context or mutual awareness, trust becomes brittle.

It is a bet without reciprocity. The system does not know that we are afraid. It cannot perceive urgency. It cannot share in uncertainty.

This asymmetry matters. Patients place trust in systems that do not understand—or even perceive—the act of being trusted.[9] That trust is fragile because it is not relational. It is one-sided.

To trust AI is to accept guidance from a system we may not fully understand. To let it shape decisions—perhaps without realizing it. A national survey conducted in Canada found that even healthcare professionals and well-informed users expressed discomfort with AI, citing concerns about transparency and control.[10] Familiarity does not always lead to comfort.

This discomfort is not limited to the general public. A recent systematic review found that more than 60% of healthcare professionals reported hesitation to use AI tools, citing transparency gaps and concerns over data security.[11] Even among those trained in clinical practice, trust in AI remains fragile when systems are opaque, poorly governed, or insufficiently explained.

The more AI augments care—supporting a diagnosis, flagging a risk, suggesting a therapy—the more we are exposed not only to its insights, but to its assumptions. We are still vulnerable. But now, the recipient of that vulnerability cannot recognize it.

That vulnerability extends into how we train future clinicians. In a recent survey of medical and health profession students across four countries in the Middle East and North Africa, the majority reported low AI knowledge and discomfort with its growing role in care—despite being broadly supportive of its potential.[12] While most favored integrating AI into healthcare education, only a third had received formal instruction. Even enthusiasm, it seems, is not enough to bridge the trust gap.

## 4. ACCOUNTABILITY AND MORAL RESPONSIBILITY

One reason trust has remained so resilient in human healthcare—despite uncertainty and risk—is that someone is accountable. If something goes wrong, there is a person who carries the ethical and professional weight of that outcome.

Clinicians are not infallible, but they are trained to take responsibility for their decisions. That expectation is built into the very culture of medicine, reinforced by codes of conduct, licensure, and legal oversight.

There is comfort in knowing that someone stands behind the decision—not just because they made it, but because they will answer for it.

When we speak with a clinician, we know who we are speaking to. We can ask them to explain their reasoning. We can challenge their thinking. We trust not just their training, but the moral framework that comes with their role. They have taken an oath—and we believe they mean it.

That sense of moral agency matters. It makes it possible to place trust in someone we barely know, because we trust the structure around them: the profession, the standards, the expectations of accountability.

Yet as AI enters the picture, that clarity begins to shift. AI tools increasingly assist in diagnosis, triage, treatment planning, and risk prediction. And while responsibility formally still rests with the clinician, the reality is often less clear.

If a tool recommends a course of action and the clinician accepts it, who is truly responsible for that decision?

When a recommendation comes from another human, we can ask for their reasoning. When it comes from a model, the logic may be opaque—rooted in training data, algorithmic structure, or parameters that even the developer can't fully explain.

The role of the radiologist is also evolving, and fast. Once defined mainly by image interpretation, radiology is becoming a data-driven discipline—one that integrates imaging, pathology, genomics, laboratory results, clinical history, and AI insights to deliver patient-specific diagnoses. The radiologist is no longer just a reader of scans, but a diagnostic strategist, data integrator, and clinical collaborator—working at the intersection of precision medicine, digital infrastructure, and clinical care.

This expanded role demands more than clinical acumen. It calls for AI literacy, systems-level thinking, ethical awareness, and leadership across multidisciplinary teams. Radiologists are now expected to guide the sequencing of diagnostic tests, participate in tumor boards, support AI validation, and communicate findings clearly—to both colleagues and patients.[13] They must understand how AI tools are trained, recognize their limitations, and advocate for equitable, transparent, and safe diagnostic ecosystems.

In this context, trust isn't just about believing in the system's output. It's about believing that radiologists can govern that system—curating data, shaping algorithms, overseeing implementation, and guiding decision-making in high-stakes environments. When radiologists act as stewards of precision diagnostics, they don't just use AI—they help ensure that trust in it is not blind, but responsibly earned.

That stewardship role reinforces a broader truth: AI cannot be trusted in isolation. It must be embedded in professional ecosystems—anchored by human judgment, institutional accountability, and clear ethical standards.

Still, even with radiologists stepping into governance roles, frontline clinicians are often expected to trust AI recommendations—especially when these tools are deeply embedded in workflow.

The model doesn't act alone—but it strongly shapes how clinicians think. In practice, this creates a kind of soft delegation. The AI system does not replace the clinician, but it steers their attention, narrows their options, and sometimes nudges their decision-making.[14]

As AI becomes more advanced, responsibility may no longer be as easily assigned. If an AI model makes a mistake—but the human follows it—who is to blame?

And what happens when models update themselves, making subtle changes with little notice?

This raises uncomfortable but necessary questions. Will clinicians always feel empowered to override AI advice, or will time pressures and workflow dependencies make deference the path of least resistance?

As automation becomes more seamless, responsibility may begin to blur—not because people abandon it, but because the system no longer requires them to claim it.

This is not theoretical. A growing body of research warns that the clinical environment often lacks the structural clarity to support shared responsibility in AI-enabled settings.[14]

Responsibility becomes diffused—split across developers, vendors, institutions, and clinicians—each essential, yet none clearly accountable. Without strong clinical leadership—such as radiologists overseeing tool use and validation—there's often no clear point of moral or legal accountability. The result is a system where trust erodes not because people act in bad faith, but because no one is clearly in charge.

To address this, some argue that trust must shift from the AI itself to the larger sociotechnical system that surrounds it.[15] AI systems, they note, lack emotion, intent, or moral awareness—so they cannot be held accountable in the way humans can. But trust can still be earned if accountability is embedded into the system across multiple layers: through transparent design, institutional oversight, and enforceable standards of responsibility.

Emerging governance models like the "Trust Octagon" reflect this view.[16] Rather than relying on abstract ethical principles, they outline specific, assessable dimensions of trustworthiness—including robustness, fairness, transparency, legal compliance, and social responsibility.

The goal is not to humanize AI, but to make it governable.

Trust in AI is not just about outcomes. It is about ensuring that responsibility is never ambiguous, and that someone—somewhere—will stand behind the decisions being made.

## 5. TRANSPARENCY AND EXPLAINABILITY

When we trust a person, we don't expect perfect knowledge. We don't need to understand every detail of a surgical procedure to trust a surgeon. But we do need clarity—the ability to ask questions and receive answers that make sense. We need reasoning we can engage with, and decisions we can challenge.

This kind of transparency sustains trust in healthcare. It's not about complete information, but about enough understanding to participate meaningfully in our own care. In fact, research shows that patients' willingness to seek care is shaped less by full understanding than by whether they trust the system's intent and explanations.[6] This kind of trust predicts action—even when information is incomplete.

Clinicians can provide that kind of explanation. Even in uncertainty, they can walk us through how they weigh symptoms, risks, and history. Their reasoning is not always complete, but it is accessible. We may not follow the entire chain of logic—but we know where to start asking.

AI systems make that harder. Their recommendations may be accurate, but they often operate on patterns and abstractions that exceed human reasoning—leaving even developers struggling to explain why a result was flagged. And that's where trust begins to fray. A system that performs well but feels opaque is hard to rely on, especially in moments of uncertainty.

And this isn't just theoretical discomfort. A recent review found that over 60% of healthcare professionals hesitate to adopt AI systems due to persistent gaps in transparency and data security.[11]

It's not enough for AI to be right. It must also be accountable to reason. In one study, people trusted AI more after it made a mistake—if they were offered a plausible rationale for why it failed.[17] Without explanation, errors became trust-breakers. With it, they became forgivable.

Yet explainability is no guarantee of confidence. In a recent clinical study, even when AI systems provided detailed visual explanations, they did not increase trust or satisfaction. For some users, explanations actually heightened uncertainty.[8]

This reflects a deeper tradeoff. The most powerful AI systems—those that outperform humans at specific tasks—are often the least explainable. Simpler models offer transparency, but not always performance. Complex models offer results, but little insight.[18]

What users often want is not a feature map—but a signal of confidence. In fact, some clinicians report they would rather see uncertainty estimates than internal logic.[19]

Trust, in these settings, is often calibrated not through deep understanding, but through visible signals of reliability—such as confidence bands, error ranges, or acknowledgments of uncertainty.[8] These can make AI feel less like a black box and more like a partner clinicians can work with.

In practice, trust is not about technical transparency. It's about shared understanding. We need to know enough about the system—how it was trained, what its limits are, and when to be cautious.[5] We don't need full interpretability, but we do need insight into what kind of system we're working with—and whether it aligns with how we think and care.

This is where system-level approaches become essential. Frameworks like the Trust Octagon offer practical criteria—fairness, robustness, transparency, and legal safeguards—that help institutions evaluate trust not just in tools, but in how those tools are deployed.[16]

Others argue that explainability alone is not enough. Trust also hinges on privacy, accountability, and perceived fairness—especially in high-stakes care.[15] These values must be embedded in governance, not just interfaces.

Even when users don't fully understand how AI works, they often respond to design cues that feel human—like personalized outputs or conversational tone.[3] This kind of emotional trust can shape behavior more strongly than accuracy alone.

And even familiarity doesn't guarantee comfort. In a Canadian survey, many respondents—especially clinicians and middle-aged women—expressed discomfort with AI despite high familiarity.[10] Trust, it seems, is not just informational. It's relational.

Because trust is not about seeing every gear inside the machine. It's about knowing when the machine should be running—and when to stop it.

## 6. CONTINUITY, CHANGE, AND ADAPTIVE TRUST

One reason people continue to trust healthcare—despite its flaws—is because it moves carefully. Medicine evolves slowly. Guidelines change after years of deliberation.

Treatments are introduced with rigorous oversight. Even clinical rituals—from how rounds are done to how charts are written—reinforce a sense of continuity.

That slowness signals safety. It tells us the system is cautious, measured, and serious about consequences. It builds trust not just in outcomes, but in the process behind them.

AI does not move that way. AI learns fast, updates quickly, and often shifts invisibly. Models can improve in the background—retraining on new data, re-weighting variables, refining predictions. When that happens without oversight or explanation, trust begins to falter. What worked yesterday may not behave the same way tomorrow.

The introduction of continually learning AI systems magnifies this challenge. These models update in real time, adjusting to each new data point. That creates value—but also volatility. It blurs the boundary between testing and deployment. And it makes validation harder. How do we evaluate a model that's always changing?

Traditional regulatory models are not built for this kind of evolution. But trust depends on clarity: on knowing when a system is stable, when it has changed, and how those changes affect decisions.

To keep pace, trust in AI must become adaptive. We can no longer treat trust as a one-time commitment. It must be a process—a belief that updates over time, based on performance, transparency, and alignment with care values.[20]

That means clinicians need tools that help them track how AI models evolve. They need to know when recommendations shift, how often updates occur, and what data is driving those changes.

But even with good tools, trust can falter if the broader system isn't ready. Deploying AI in misaligned workflows—fragmented data, poor incentives, limited oversight—undermines adoption and weakens trust.[11] A shift in infrastructure and mindset is needed to support adaptive trust at scale.[11]

Some systems now include dashboards for evaluating AI trustworthiness in real time—assessing consistency, reliability, and perceived value.[21] But most clinical settings still lack those mechanisms.

Even tools built for transparency sometimes miss the mark. In one study, clinicians didn't want more detailed explanations—they wanted to see how uncertain the model was.[8] That small shift reframes what it means to be transparent: not full interpretability, but shared situational awareness.

And without that awareness, trust becomes brittle. A recent review of over 50 studies found that clinicians' trust in medical AI is shaped by factors like usability, explainability, education, and perceived fairness.[19] Continual updates put pressure on all of those.

If a system changes silently, trust doesn't adapt—it erodes.

We need new ways to monitor AI as it learns. That means built-in auditing, ongoing validation, and transparent communication of change. Just as medicine monitors the body's condition over time, AI must be monitored for performance drift and safety risk.

Because trust is not just a product of accuracy. It's a response to continuity—and a test of how we manage change.

## 7. HUMAN-LIKENESS AND OTHERNESS

Much of the trust we place in healthcare is shaped by human connection—the warmth of a voice, the reassurance in someone's eyes, the quiet signals that say, I see you. These moments matter. They turn treatment into care. They make vulnerability feel safe.

AI offers none of that. It doesn't recognize pain. It doesn't share uncertainty. It doesn't respond to fear with empathy. Even the most advanced AI systems don't understand presence. They simulate engagement—but don't feel it. They generate care plans—but don't care.

Still, we're being asked to trust them.

When we trust people, we often trust what we recognize. We assume others understand pain because they have felt it. That shared humanity makes it easier to hand over something fragile—our health, our worry, our hope. It's not just about competence. It's about likeness.

That's what AI lacks. It's not someone. It's something.

And that difference matters. Studies show we're more likely to trust a humanoid robot—especially one that makes a verbal promise—if we perceive it as human-like.[22] But that trust can be misleading. When systems pretend to understand us, without actually doing so, the result isn't comfort. It's confusion.

Simulated empathy isn't the same as real care. In fact, it can feel like a lie.

Rather than trying to make AI more human, we should focus on making it more trustworthy—through design, governance, and performance. As others have argued, AI should not be the object of trust on its own. It should be part of a system that earns trust through clarity, accountability, and alignment with care values.[12]

That means moving away from imitation and toward integrity. Not trying to replicate human intuition—but supporting human expertise. Not simulating bedside manner—but reinforcing bedside judgment.

Because AI doesn't need to feel human to be trustworthy. It just needs to help humans care better.

## 8. CONCLUSION

Trust is not a thing to be transferred. It is a relationship to be redefined.

AI is changing medicine—but trust in AI will never be the same as trust in a human caregiver. One is emotional and relational. The other is technical and structural. One connects to our story. The other connects to our data. And yet both are now part of how care is delivered.

In radiology especially, where AI already shapes decisions across the diagnostic chain—from acquisition to reporting to communication—the challenge is not just to use these tools, but to govern them.

The question isn't whether we should trust AI. It's whether AI is designed in a way that deserves trust.

That means moving beyond vague promises of "trustworthy AI" toward systems that earn confidence through accountability, transparency, and alignment with the values that define good care. AI should not replicate human intuition. It should strengthen human judgment. It should not replace connection. It should support it.

Trust in AI should not be demanded—or assumed. It must be earned.

And it can be—if we design with care, implement with humility, and govern with clarity.

Because in the moments that matter most, trust is not about whether AI can think or feel. It is about whether it helps us care for one another better.

## ACKNOWLEDGMENT

During the preparation of this work, the author used OpenAI ChatGPT4o to help identify and correct grammar and typographical errors. The author reviewed and edited the content as needed and takes full responsibility for the final publication.